\documentclass[%
 reprint,
 superscriptaddress,
 amsmath,amssymb,
 aps,
 floatfix,
]{revtex4-2}

\usepackage{graphicx}
\usepackage{dcolumn}
\usepackage{bm}
\usepackage[hidelinks]{hyperref}
\usepackage[mathlines]{lineno}

\begin{document}

\title{Inverse design of plasma metamaterial devices with realistic elements}

\author{Jesse A. Rodr\'iguez}
\email{jrodrig@stanford.edu}
\affiliation{Department of Mechanical Engineering, Stanford University, Stanford, CA 94305}
\author{Mark A. Cappelli}
\affiliation{Department of Mechanical Engineering, Stanford University, Stanford, CA 94305}

\date{\today}

\begin{abstract}
In an expansion of a previous study \cite{RodriguezPRA}, we apply inverse design methods to produce two-dimensional plasma metamaterial devices with realistic plasma elements which incorporate quartz envelopes, collisionality (loss), non-uniform density profiles, and resistance to experimental error/perturbation. Finite difference frequency domain simulations are used along with forward-mode differentiation to design waveguides and demultiplexers operating under the transverse magnetic polarization. Optimal devices with realistic elements are compared to previous devices with idealized elements, and several parameter initialization schemes for the optimization algorithm are explored, yielding a robust procedure for producing such devices. Demultiplexing and waveguiding are demonstrated for microwave-regime devices composed of plasma elements with reasonable space-averaged plasma frequencies $\sim10$ GHz and a collision frequency $\sim1$ GHz, allowing for future \textit{in-situ} training and experimental realization of these designs.
\end{abstract}

                    
\maketitle

\section{Introduction}
Inverse design is a technique used to optimize device parameters given an \textit{a priori} set of design criteria and metrics. When applied to electromagnetically-active systems \cite{minkov2020inverse, ceviche, Su2020Spins, burger2004inverse, BorelTopologyWaveguide, LinPhotonicDirac, miller2013photonic,  ChristiansenTunableLens, PestourieMetasurfacesDemultiplex, ChungTunableMetasurface, MeemInverseFlatLens, piggott2015inversebinarized, liu2013transformation, hughes2018adjoint, molesky2018inverse, christiansen2021inverse,  designsurvey, AndradeInverseGuide}, devices are conceived by solving Maxwell's equations for a particular material domain and set of sources, the properties of which (e.g., relative permittivity, permeability, source frequency and power) are mapped to trainable parameters. These parameters can be modified to alter the propagation of the source fields through the device in a manner which fulfills the performance criteria. Typically, these performance criteria are encoded as an objective function that depends on the input parameters which is minimized or maximized via the inverse design algorithm. As an example, in many scenarios, the objective function involves an inner product between the desired electromagnetic fields and those obtained via simulation in regions of interest. Automatic differentiation allows the calculation of a numerical gradient of such an objective with respect to the parameters that encode the permittivity structure of the training region. The inverse design algorithm then iteratively adjusts these parameters to minimize the distance between the simulated fields and the desired field in the chosen inner product space. The result is a novel, highly specialized device that meets the design criteria of the user\cite{minkov2020inverse, ceviche, Su2020Spins}.

Unfortunately, the set of optical devices which can be experimentally realized is a small subset of the set of devices which can be computationally modeled, and so inverse design often works within a constrained setup where various limitations on the device geometry are imposed. For example, allowing for a continuous range of permittivities throughout the simulation domain results in fast optimization with very effective designs, but achieving continuous variation of permittivity throughout the design region is essentially impossible using conventional manufacturing techniques. In prior work, inverse design practitioners take advantage of more constrained systems such as photonic crystal-style devices \cite{burger2004inverse, miller2013photonic,minkov2020inverse, BorelTopologyWaveguide, LinPhotonicDirac} and metasurfaces \cite{ChristiansenTunableLens, PestourieMetasurfacesDemultiplex, ChungTunableMetasurface, MeemInverseFlatLens} where a simple map from the parameterized design region to a physically realizable device is guaranteed to exist. These maps typically encode restrictions on the range and spatial distribution of the domain permittivity (i.e. using only permittivity values and geometry corresponding to available materials and manufacturing processes). A common technique in the literature is the use of nonlinear projection to design binarized photonic devices \cite{piggott2015inversebinarized} which can be printed or created via lithography as a single-use design. Each method for constraining inverse design algorithms introduces its own limitations to the device capabilities, but even simplistic devices can be optimized to perform tasks which would normally require expert-level human design to achieve the same or worse performance.

\begin{figure}[htbp]
\centering
\includegraphics[width=\linewidth]{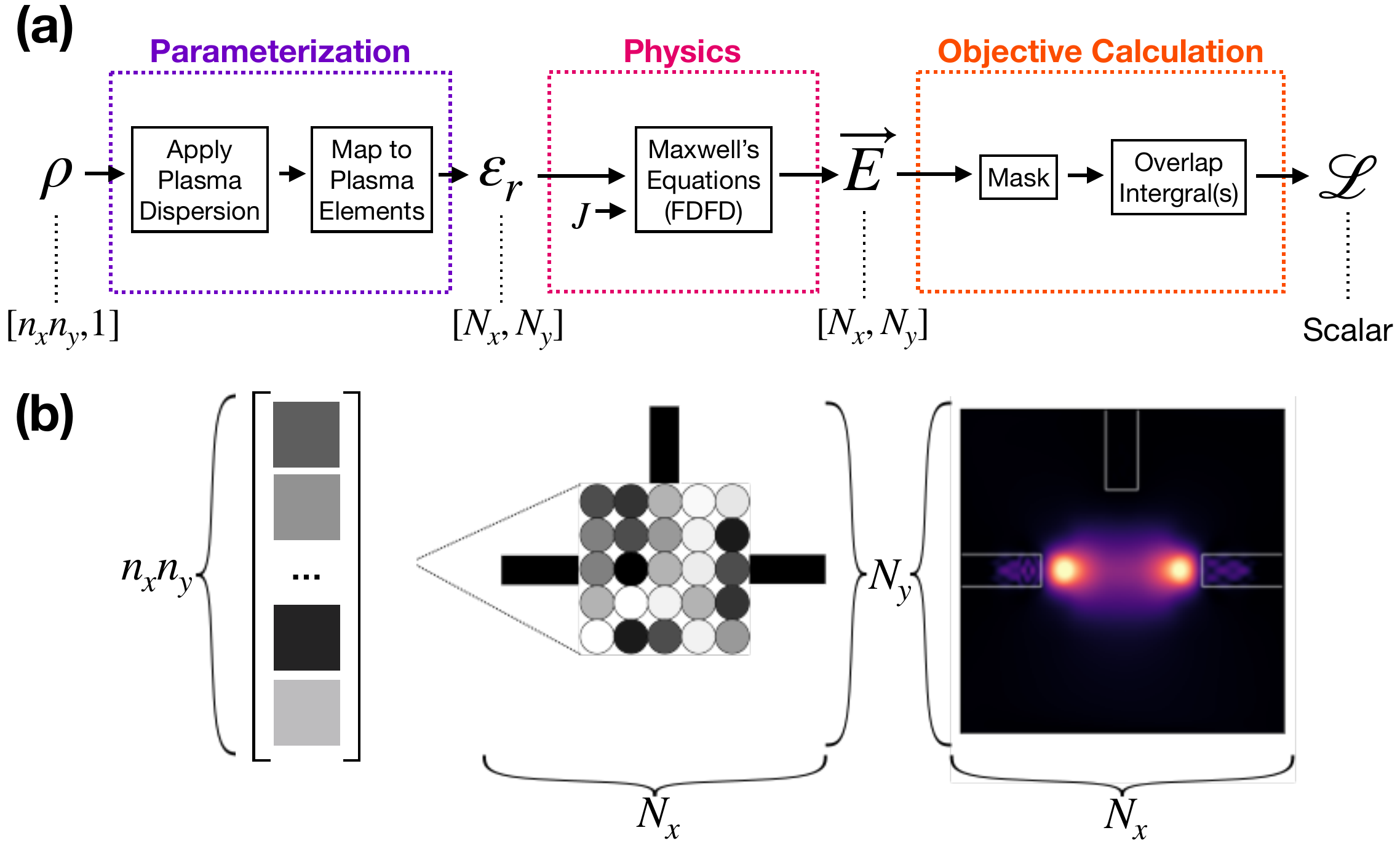}
\caption{(a) Flow chart describing the algorithmic design of our PMM array. $J$ represents the modal source for the FDFD simulation. Examples of the training parameter vector $\rho$, permittivity matrix $\varepsilon_r$ and simulated $\textbf{E}$ fields for a simple PMM device are provided in (b).}
\label{fig:invdeschart}
\end{figure}

Rather than restricting permittivity values within our domain to some discrete set, we focus instead on limiting the spatial configuration by parameterizing the design region as a plasma metamaterial (PMM). Our PMM is a square 10-by-10 array of cylindrical gaseous plasma elements (rods) which can be experimentally realized via the use of discharge lamps or laser-generated plasmas. Because plasma density (and therefore permittivity) in sources such as these can be continuously tuned through a wide range of values, our PMM serves as a good candidate for inverse design as the continuous variability of the permittivity of the elements yields an infinite configuration space for training purposes \cite{configspaceppc} and, perhaps more importantly, allows a single device to serve multiple functions. By changing the discharge current or laser power supplied to each of the plasma elements within the device, one could quickly switch between different device configurations. Previous work with inverse design electromagnetic devices achieved reconfigurability via methods such as refractive index changes \cite{ChristiansenTunableLens} in materials like liquid crystals \cite{ChungTunableMetasurface}, although the use of our PMM configuration is particularly unique as completely disparate functions can be performed with a single device as can be seen in our prior study where the PMM could function as both a demultiplexer and a photonic logic gate \cite{RodriguezPRA}. The inverse design process for our PMM geometry is summarized in Fig. \ref{fig:invdeschart}(a). In our algorithm, there is a one-to-one correspondence between each plasma element and the $n_xn_y$ elements of the parameter vector $\rho$ (Fig. \ref{fig:invdeschart}(b)). In the end, the parameterization scheme results in a relative permittivity matrix $\varepsilon_r$ (of size $N_x\times N_y$, dependent on the simulation resolution) containing permittivity values at each pixel in the simulation domain. 

In addition to the complex refractive behavior which can be accessed in PMM devices, when the PMM is composed of elements that are small compared to the operating wavelength and the source is polarized properly, we can couple into localized surface plasmon (LSP) modes along the boundary between a positive permittivity background and a negative permittivity plasma element. Such surface modes can yield more complex and efficient device behavior \cite{tunableppc, SakaiNPerm, SakaiChain}, allowing for a particularly high degree of reconfigurability, but for devices which seek to maximize transmission, they can cause serious degradation of performance \cite{RodriguezPRA}. Prior experimental studies of plasma photonic crystals (PPCs) have already highlighted the richness of this geometry in waveguide \cite{RodriguezPRA, guideppc} and bandgap \cite{tunableppc, ppc, tunablepla} devices. The complicated array of electromagnetic phenomena present in PMM devices is well-matched to inverse design methods where more subtle physical phenomena can be leveraged by the optimization algorithm in a manner that would be impossible in a human-designed device.

\section{Methods}
\begin{figure}[htbp]
\centering
\includegraphics[width=0.85\linewidth]{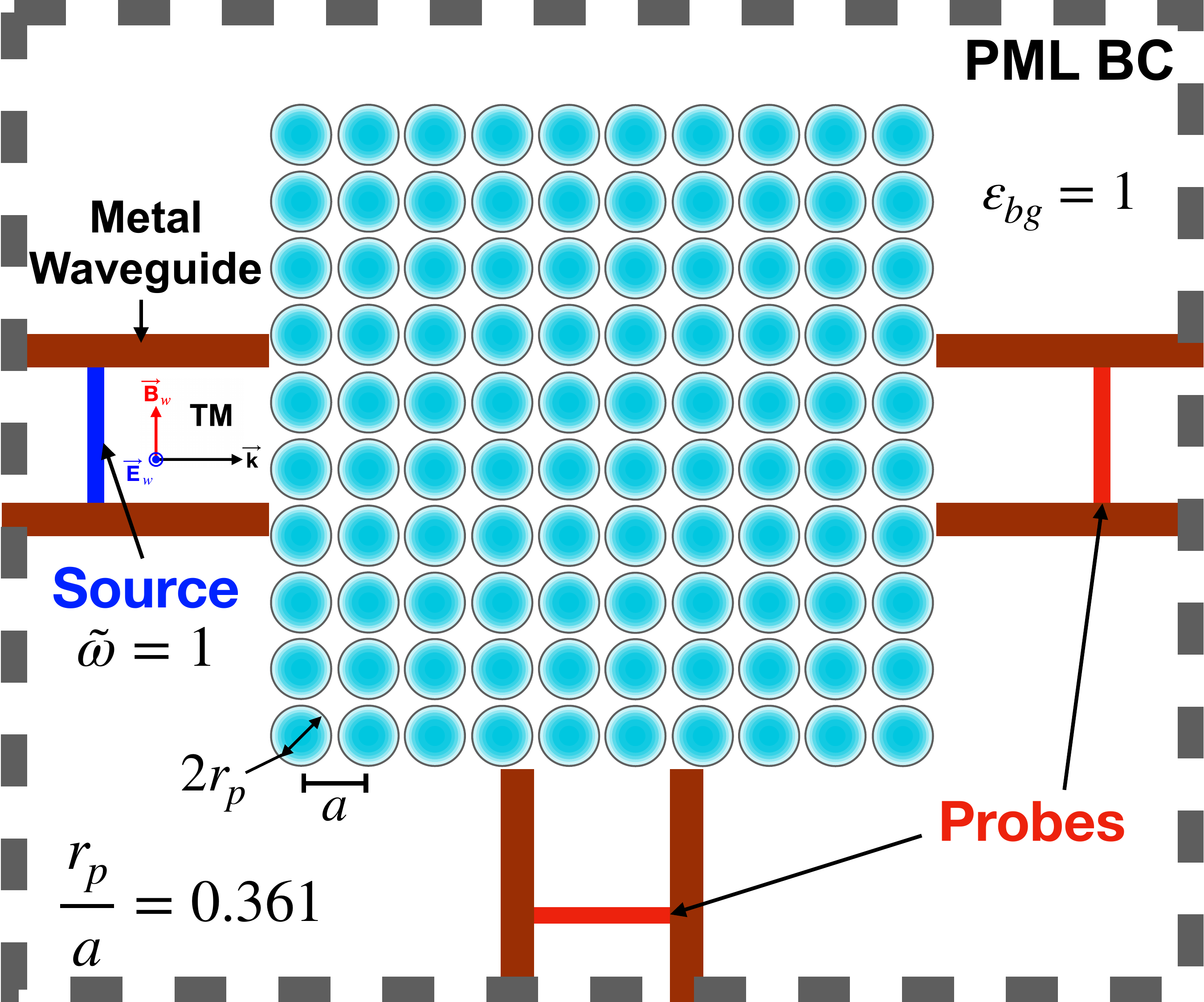}
\caption{Schematic diagram for the PMM device in a waveguide configuration. The device consists of a $10\times10$ array of plasma discharge bulbs (cyan) suspended in air with metal waveguides (brown) functioning as the entrance and exit(s). The 'probe' slices (red) represent the domain over which the $L_2$ inner products that constitute the objective function are calculated. $\varepsilon_{bg}$ is the background permittivity. $\tilde{\omega}$ is the nondimensionalized operating frequency $\omega/(2\pi c/a)$.}
\label{fig:wvgschemchart}
\end{figure}

Fig. \ref{fig:wvgschemchart} provides a schematic of our PMM device in a directional waveguide configuration. A demultiplexer configuration, described later in the paper, utilizes the exact same array of tunable plasma elements but with different entrance/exit waveguides. The device is simply a 10$\times$10 array of tunable plasma discharge tubes suspended in air and spaced according to the limits of existing experimental facilites \cite{guideppc,ppc}. A detailed description of the plasma elements is found below. We chose a 10$\times$10 array for this study for two reasons; first, this device size is about as large as we can reasonably produce experimentally in the near future, and second, stepping up to a larger array severely limits our ability to train devices in a reasonable amount of time. Although the raw simulation time simply scales as the square of the device dimension since the simulations are two-dimensional, the current design of the optimization scheme requires a large amount of memory, and so going any larger than 10$\times$10 would necessitate the use of shared high-performance computing (HPC) resources for device training rather than our lab's own resources; requiring long wait times that are not conducive to rapid readjustments to the training parameters. We do believe that larger arrays would enable the device to perform more complex functions, but this will be left to future investigations.

The PMM device functions by redirecting modal sources introduced at the entrance waveguide(s) to the desired output waveguide(s). In this work, the electromagnetic field solutions are calculated via finite difference frequency domain (FDFD) simulations computed with Ceviche, an autograd-compliant electromagnetic simulation tool \cite{ceviche} that enables calculation of the gradient of the optimization objective with respect to the input parameters that encode the simulation domain. Note that the use of FDFD implies that all computed devices represent the steady state solution achieved after some characteristic time. The simulation domain was discretized using a uniform square mesh with a resolution of 75 pixels per lattice constant $a$ due to the CPU memory and runtime limitations mentioned above, resulting in a domain size of $N_x=N_y=1500$ pixels in each case. The 'optimal' devices were then run at higher resolutions up to 150 pixels per lattice constant on a HPC cluster to confirm that the device functionality was preserved; ensuring that the training resolution was high enough. A perfectly-matched boundary layer (PML) $2a$ in width was applied along the domain boundary. 

The polarization of the input source has a strong effect in 2D devices such as these, either \textbf{$E_z$} ($E$ out of the page), which we call the TM polarization, or \textbf{$E_y$} ($H$ out of the page), which we call the TE polarization. Prior work with PPCs indicates that the response of the device to both TE and TM-polarized sources is highly tunable, with the former case benefiting (and suffering, depending on the desired device functionality) from the presence of LSP modes \cite{righetti2018enhanced, tunableppc, SakaiNPerm, SakaiChain}, while the latter makes more direct use of dispersive and refractive effects \cite{tunableppc, guideppc, ppc}. In our previous study \cite{RodriguezPRA}, we see that for devices which seek to preserve an input signal like the waveguides and demultiplexers we present below, the TE polarization leads to poor performance, so we choose to focus on the TM mode alone in this study. Thus, we will not expect to see any small, sub-wavelength-scale field structures around our plasma elements or the large transmission losses which are indicative of the presence of LSP modes.

The simulated $\mathbf{E}$ fields are masked to compute the $L_2$-inner product integrals along planes of interest within the problem geometry. These integrals make up the optimization objective $\mathcal{L}(\rho)$. Our simulation tool, Ceviche, is then used to compute the numerical gradient of the objective function with respect to the training parameters via forward-mode differentiation \cite{ceviche}. Ceviche uses the Adam optimization algorithm \cite{kingma2014adam} (gradient ascent with momentum, in essence) to iteratively adjust $\rho$ and thereby maximize $\mathcal{L}$, or rather, minimize the $L_2$ distance between the simulated fields at the device output and the desired fields. Optimization was conducted with learning rates ranging from $0.0002-0.01$, and with typical values for the Adam hyperparameters; $\beta_1=0.9$, and $\beta_2=0.999$.

To summarize, the PMM optimization problem can be expressed as
\begin{align*}
    &\max_{\rho} \quad  \mathcal{L}(\mathbf{E}) \\
    &given \quad \nabla\times\frac{1}{\mu_0}\nabla\times \mathbf{E}-\omega^2\varepsilon(\omega,\rho)\mathbf{E}=-i\omega J.
\end{align*}
where $\mathcal{L}$ is the objective composed of a set of $L_2$ inner product integrals of the simulated field with the desired propagation mode, $\textbf{E}$ is the electric field, $\rho$ is an $n_xn_y$-dimensional vector that contains the parameters that set the electron density of each of the PMM elements, $\mu_0$ is the vacuum permeability, $\omega$ is the field frequency, and $\varepsilon(\omega,\rho)$ is the spatially-dependent permittivity that is encoded by $\rho$, and $J$ is a current density used to define a fundamental modal source at the input waveguide. In practice, the permittivity distribution among the plasma elements is controlled by varying the discharge current (which therein alters the plasma density) in each of the PMM elements according to the Drude model, which is where the dependence on $\omega$ arises. The non-dimensionalized plasma frequencies in $\rho$ are mapped to element permittivities via the Drude model;
\begin{align*}
    \varepsilon=1-\frac{\omega_p^2}{\omega^2+i\omega\gamma}
\end{align*}
where $\gamma$ is the collision/damping rate, $\omega_p^2=\frac{n_ee^2}{\varepsilon_0m_e}$ is the plasma frequency squared, $n_e$ is the electron density, $e$ is the electron charge, $m_e$ is the electron mass, and $\varepsilon_0$ is the free-space permittivity. 

The devices that follow have been explored previously in ref. \cite{RodriguezPRA}, but mostly as a proof of concept with very simplistic plasma elements that were collisionless, uniform-density rods with no quartz envelope. In this study, we attempt to model all major non-ideal factors in order to assess whether or not such devices will be readily experimentally realized. Real gas discharge plasmas will be affected by a range of non-ideal factors in addition to the damping rate included in the Drude model above. 

\begin{figure}[htbp]
\centering
\includegraphics[width=0.85\linewidth]{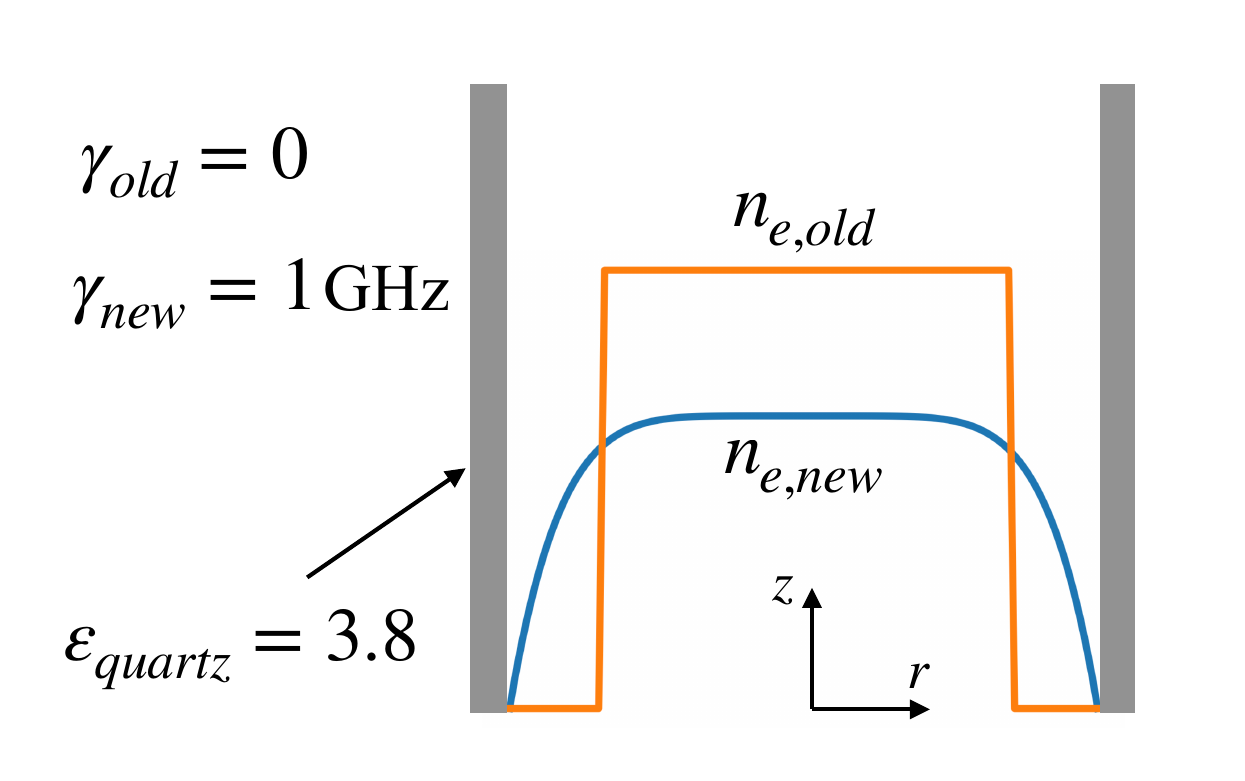}
\caption{Illustration comparing our model (labeled 'new') and the idealized model (labeled 'old') from ref. \cite{RodriguezPRA} for the plasma discharge tubes. The idealized model does not include the quartz bulb.}
\label{fig:densprof}
\end{figure}

Collisional damping in discharges can be brought to very low levels in practice by reducing the discharge pressure, but this will come at the expense of plasma density (and correspondingly, plasma frequency), which in turn restricts the range of values of the plasma dielectric constant that we can access. Furthermore, the plasma density profile within the gaseous elements is not uniform and dependent on the methods used to generate the plasma. For example, within a discharge tube at nominal operation conditions, the plasma has a core of relatively uniform density plasma and then a sheath region near the bulb where the density drops to negligible values at the inner bulb surface. As one would expect, the plasma density profile can have a strong effect on device performance in both the TE mode by affecting LSP coupling \cite{SakaiNPerm, SakaiChain} and in the TM mode by altering the refraction of impinging waves. In the past, our group has approximated this with a uniform plasma column of radius $r_i/\sqrt{2}$ where $r_i$ is the inner radius of the bulb's quartz envelope. This leads to decent agreement with experiment in photonic crystal configurations \cite{guideppc, ppc, righetti2018enhanced, tunableppc}, but may not be appropriate for simulating devices with EM wave propagation structures as complicated as are found in our inversely designed PMMs. To best account for the non-uniform density profile, we choose as our model a 6th-order polynomial as a function of radius, where the trainable parameters $\rho$ map to the space-averaged plasma frequency of each of the rods. Although a simple analysis considering diffusion and recombination at the plasma-quartz interface would suggest a Bessel function solution which would be better approximated by a 2nd-order polynomial, the 6th-order polynomial profile has been shown to lead to the closest match between simulated and observed transmission spectra for our PPC devices \cite{ThomasThesis}. In addition, we include the $\varepsilon=3.8$ quartz bulb around the plasma elements that was omitted in ref. \cite{RodriguezPRA}. Figure \ref{fig:densprof} illustrates the difference between the simplistic plasma elements in ref. \cite{RodriguezPRA} and our more realistic ones.

In addition to all of this, while an accurate and precise control of element permittivity by varying discharge current is possible in theory, in practice a number of factors (such as thermal variation, mechanical/electrical error, etc.) can lead to small unknown perturbations in discharge current $\sim 1$ - $5$\%, which leads to density perturbations of about the same order of magnitude relative to the nominal density. Thus, if one wishes to translate these designs to a physical device, such perturbations must be accounted for. This can be done in one of three ways: 1) random noise can be programmed into the learning algorithm itself at greater computational cost as training would take longer, 2) the device can be trained \textit{in-situ} which would automatically take experimental noise within the apparatus into account, or 3) one can optimize the device without taking any noise into account and then do a sensitivity test on the optimized parameters to determine how resistant the resultant designs are to perturbations in plasma density. In this study, we carry out the third option.

The manner in which we conduct the sensitivity test is as follows: Since the elements of the device would be individually controlled in practice, we add mean-zero independently distributed Gaussian noise to the density of each element with a standard deviation $\sigma=p\sqrt{n_e}$ where $p$ represents a 'perturbation factor' that sets the variance of the random noise. If $p=0.05$, we can expect variation in plasma density amongst the columns on the order of $\sim 5$\% of the plasma density in each individual column. This is the same as reassigning the plasma density of each element according to 
\begin{align*}
	n_{e,perturbed} = \left|n_{e,opt} + \mathcal{N}(0,p^2)n_{e,opt}\right|, 
\end{align*}
where the absolute value is applied so we don't end up with non-physical negative densities when the Gaussian values are spuriously large and negative. For each device, we run 10 different random perturbations with the same value for $p$ and increase $p$ in increments of 0.01 until we see a significant failure mode in at least 3 of the 10 runs. For a failure mode to be deemed 'significant', we require a drop in transmission in the correct output port of more than 10\%. For the demultiplexer, since a drop in transmission for one frequency can be accompanied with an increase in transmission in the correct output for another frequency, we require a drop in transmission in the correct output of 10\% or greater for all frequencies. The dominant failure mode amongst the 3 or more dysfunctional perturbed devices along with the average drop in transmission in the correct port across all 10 simulations is presented below alongside the original design to illustrate how robust each device is and how the device functionality will deteriorate if the plasma density is perturbed to a high enough degree.

Once the optimization algorithm converges on a device, we proceed by conducting a quantitative evaluation to complement the E-field results. To do this, we calculate the S-parameters, insertion loss, and loss to the plasma elements via the time-averaged Poynting flux through specific flux surfaces in the domain. The time-averaged Poynting vector was calculated using the phasor field amplitude results from the FDFD calculation as $\vec{S}_p=\text{Re}\{\vec{E}\times\vec{H}^*\}$. After determining the total energy flux out of the source, the energy flux out of the input waveguide into the domain is calculated to yield the insertion loss, and the net energy flux into the plasma elements is calculated to yield the loss to the plasma elements. The locations of the source and the probes where the objective is calculated are used as the ports for the calculation of the relevant S-parameters, which we define in the following way for this study:
\begin{gather*}
    S_{11} = 10\log_{10}\left[\frac{S_{tot}-\int \vec{S}_{p}\cdot \vec{dA}_{1}}{S_{tot}}\right]\\
    S_{21} = 10\log_{10}\left[\frac{\int \vec{S}_{p}\cdot \vec{dA}_{2}}{S_{tot}}\right]\\
    S_{31} = 10\log_{10}\left[\frac{\int \vec{S}_{p}\cdot \vec{dA}_{3}}{S_{tot}}\right],
\end{gather*}
where $S_{tot}$ is the total energy flux out of the source divided by 2 (i.e. the positive-$x$ directed energy flux out of the source). A diagram illustrating the location of the ports and loss evaluation surfaces is given in figure \ref{fig:Sparams}(a).

\section{Results and Discussion}
\subsection{S-Parameters}
\begin{figure}[htpb]
	\centering
	\includegraphics[width=1\linewidth]{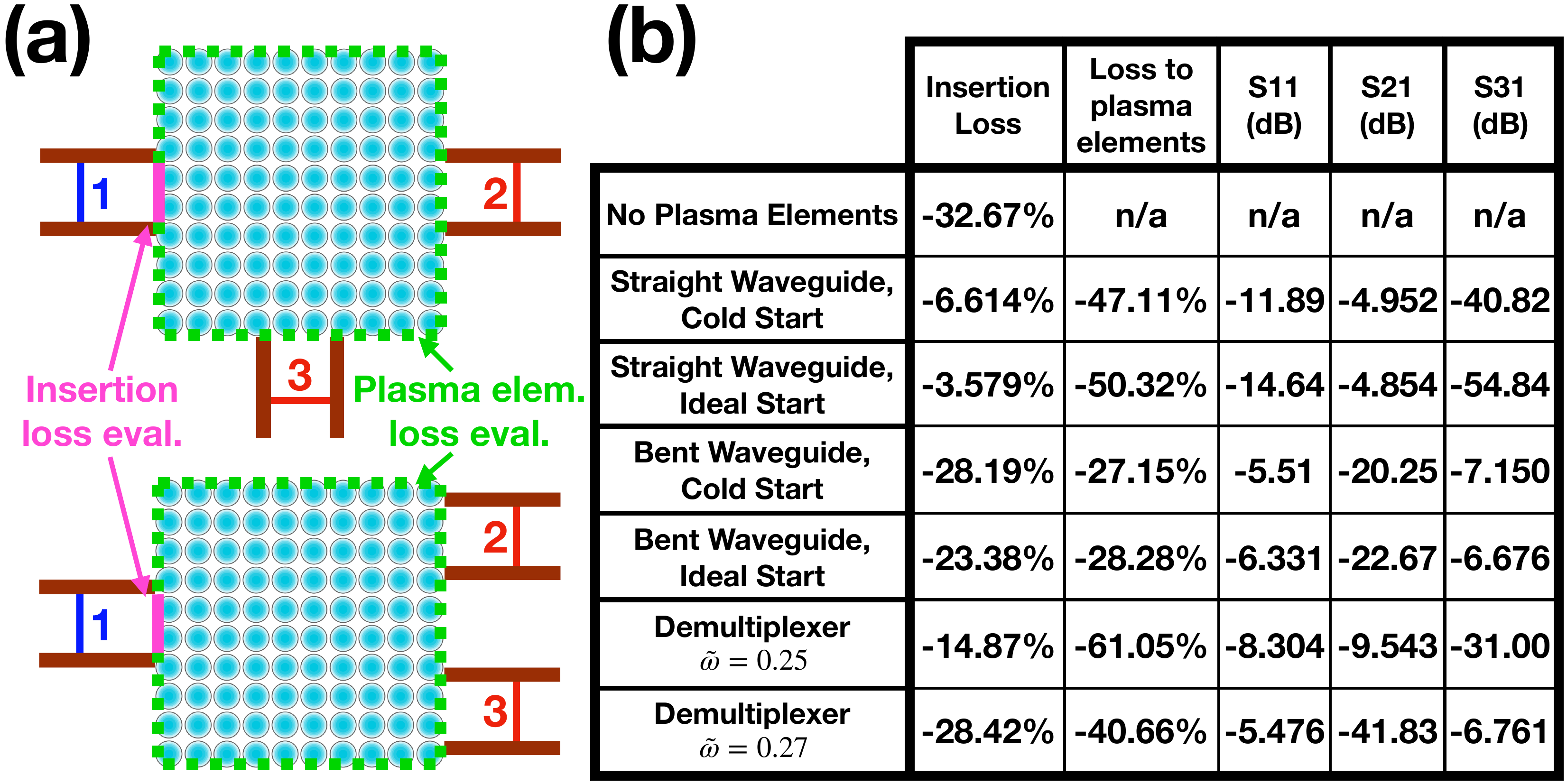}
	\caption{(a) Illustration denoting the loss-evaluation surfaces and ports for the waveguide (top) and demultiplexer (bottom) configurations and (b) Insertion loss, loss to the plasma elements, and S-parameters for each of the optimal devices presented in this study.}
	\label{fig:Sparams}
\end{figure}

In figure \ref{fig:Sparams}, the quantitative measures of performance for each of the device functions is given, with the 'cold' and 'ideal' start designations referring to whether the initial parameters before optimization were from the idealized study (ideal) or just had all the plasma elements inactive (cold). One important aspect to note before discussing each individual device is that since we didn't give any consideration to the coupling between the input and exit waveguides and the device, even when the plasma elements are not present about 1/3 of the input power is reflected at the waveguide exit interface. This insertion loss is improved in every case by the optimal design, however one can imagine that the use of a horn or some other apparatus to reduce the insertion loss would improve the overall performance of the device in all cases. As such, the relatively high S11 values in each case are not a major concern. 

In all cases, the transmission through the correct port is well over an order of magnitude higher than that of the incorrect port. The worst comparison comes in the 'cold start' bent waveguide case where there is a 13 dB difference in transmission between the two ports, and the best comes in the 'ideal start' straight waveguide case where there is a 50 dB difference in transmission between the correct and incorrect ports. In all the waveguide cases, about half of the signal power is lost due to insertion losses and dissipation within the plasma, while in the demultiplexer case these losses are a bit higher at around 70\% of the input power. Again, we stress that these losses would likely be partially mitigated by the introduction of a better coupling scheme between the inputs/outputs and the device elements. The values given in figure \ref{fig:Sparams} give us a strong quantitative evaluation of each of the PMM functions, showing that the device consistently directs much more power to the correct output than the incorrect output. 

\subsection{Directional waveguide}
\begin{figure}[htpb]
\centering
\includegraphics[width=\linewidth]{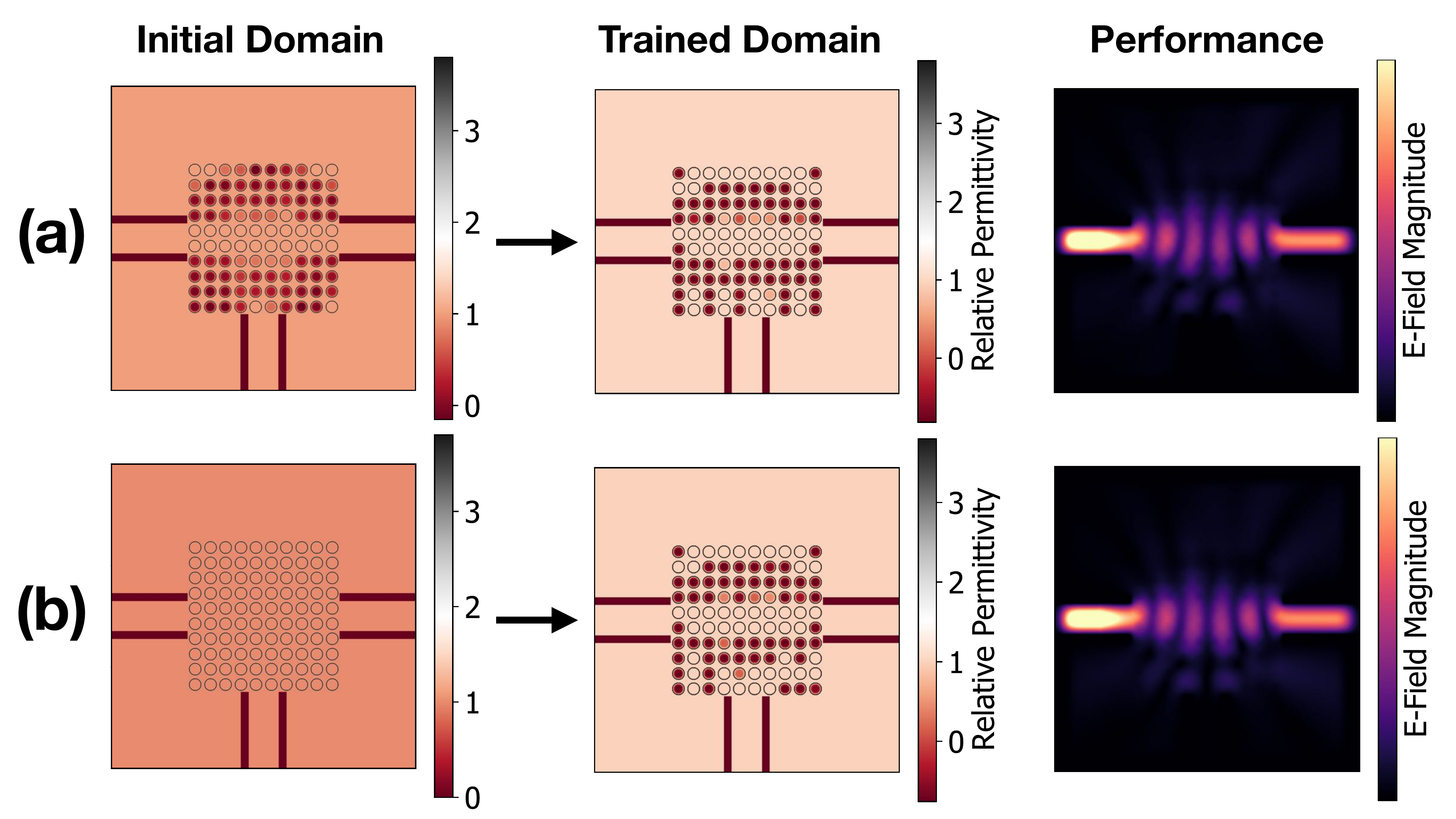}
\caption{Initial and trained relative permittivity domains and FDFD-simulated field magnitude $|\mathbf{E_z}|$ ($\hat{\mathbf{z}}$ out of the page) showing the resultant performance for the straight waveguide initialized with (a) the optimal parameters from \cite{RodriguezPRA} and (b) with all plasma elements deactivated, where the maximum plasma frequency among the plasma elements is about $\tilde{\omega}_p=0.35$ which translates to $f_p=7$ GHz.}
\label{fig:straightguides}
\end{figure}

Electromagnetic waveguides are fairly common in inverse design devices either as the primary function or as important building blocks to more complicated devices \cite{BorelTopologyWaveguide, AndradeInverseGuide}. Presented in Fig. \ref{fig:straightguides} are field simulations and initial/final (initial and final here meaning before and after training) domain permittivities of optimal straight waveguide configurations for our PMM. Operating at $\omega=0.25\times(2\pi c/a)$, or a non-dimensionalized source frequency $\tilde{\omega}=\omega/(2\pi c/a)=0.25$, this represents the simplest type of optimization problem for our PMM. Because of the simplicity of this geometry, we limit the plasma frequency to $f_p=7$ GHz, a modest operating condition that could be easily reached and run at long time scales ($\sim$ hours) without damaging the discharge bulbs. In this and all other simulations, the relative permittivity of the input and output waveguide walls was set to $\varepsilon=-1000$ to serve as a lossless metal. The waveguide objective is the $L_2$ inner product of the simulated field with an $m=1$ propagation mode (in order to preserve the input mode) evaluated at the probe in the desired exit waveguide minus the integrated field intensity at the incorrect exit to discourage leakage into the wrong output:
\begin{equation*}
\mathcal{L}_{wvg}=\int E\cdot E_{m=1}^*dl_{\text{desired exit}}-\int |E|^2dl_{\text{incorrect exit}}.
\end{equation*}
One could likely improve performance even further by discouraging all leakage; i.e. subtracting the field intensity integrated over all space outside the device other than within the correct output, but for the devices in this study the improvement is marginal. 

In the prior idealized study \cite{RodriguezPRA}, the initial parameters were set to be uniform among all the plasma elements since the initial conditions were found to be unimportant; i.e. the optimal parameters were very similar between runs with different intial parameters. Upon adding all the non-ideal factors described above, the optimization objective function became significantly more complicated with many more local minima than before, and so the initial conditions became closely linked with the device performance once the algorithm converged. In fig. \ref{fig:straightguides}, the results for both initializations look similar in both optimal parameters and field distributions, but when one considers the S-parameters of the devices given in fig. \ref{fig:Sparams} we can see that when initializing the domain with all the plasma elements turned off (fig. \ref{fig:straightguides}(b), the 'cold' start), the functionality of the device suffers, whereas if the device is initialized with the optimal parameters from ref. \cite{RodriguezPRA} (fig. \ref{fig:straightguides}(a), the 'ideal' start), the final device performs much better even though the parameters change significantly during training. The trained domain for the naively initialized device, while very similar to the device initialized with the idealized results, lacks activation of several bulbs near the incorrect exit, resulting in a signal in that port more than an order of magnitude higher than the 'ideal' start case. The 'ideal' case also exhibits much better coupling into the device from the waveguide, reducing S11.

\begin{figure}[htpb]
\centering
\includegraphics[width=\linewidth]{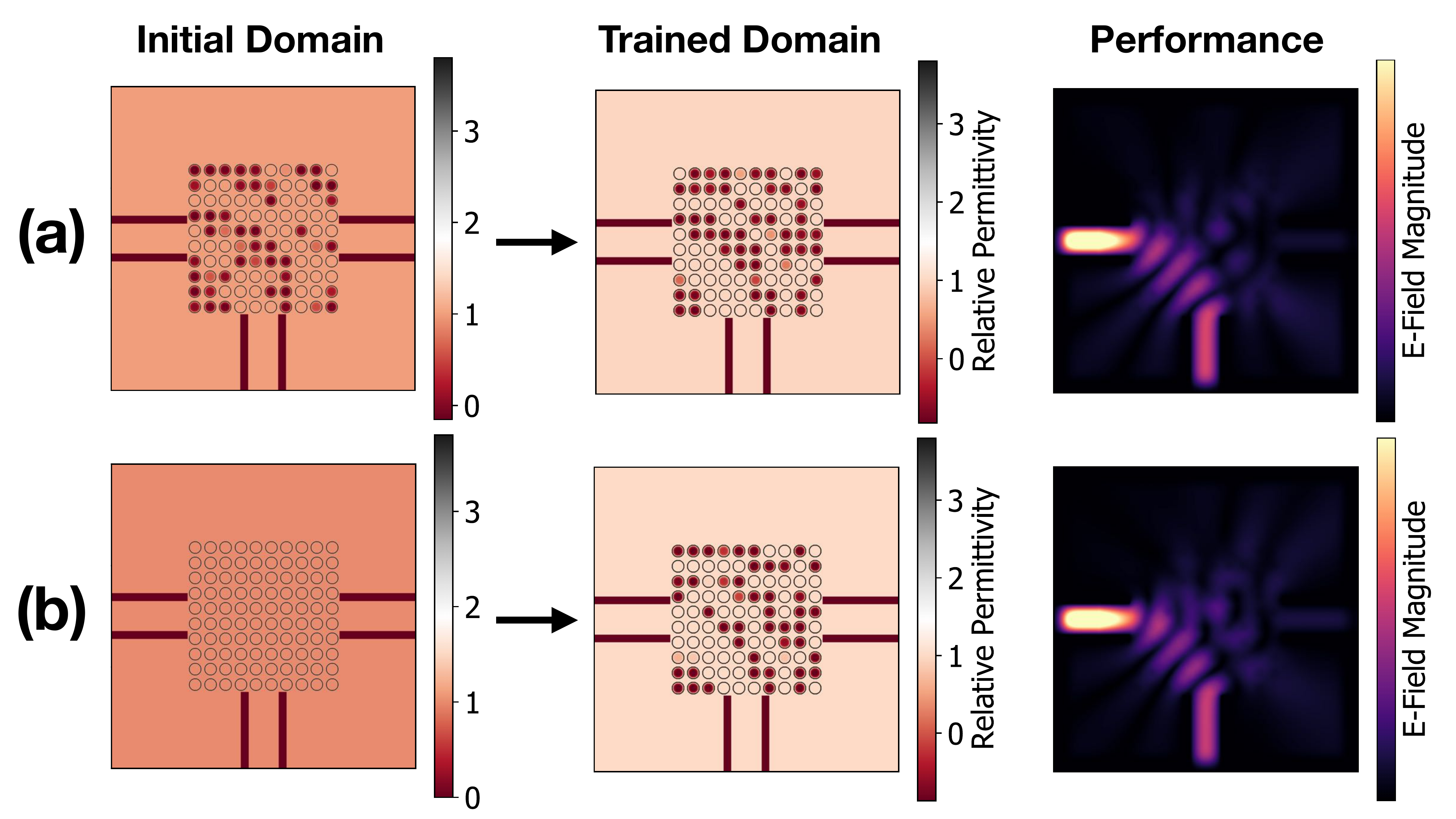}
\caption{(a) Initial and trained relative permittivity domains and FDFD simulated field magnitude $|\mathbf{E_z}|$ ($\hat{\mathbf{z}}$ out of the page) showing the resultant performance for the bent waveguide initialized with (a) the optimal parameters from \cite{RodriguezPRA} and (b) with all plasma elements deactivated, where the maximum plasma frequency among the plasma elements is about $\tilde{\omega}_p=0.35$ which translates to $f_p=7$ GHz.}
\label{fig:bentguides}
\end{figure}

Next, we modified the objective to produce a waveguide with a 90-degree bend by simply switching the role of the desired and undesired exit probes in the straight waveguide objective. The results for this case are found in Fig. \ref{fig:bentguides} where the difference in performance between the two cases is slightly easier to observe in the field plots. Again we see that, while both cases result in decent performance, the device initialized with the optimal parameters from ref. \cite{RodriguezPRA} performs better with less loss to the surroundings and lower insertion loss. As this objective is obviously more difficult than the straight waveguide, there is a much smaller difference in transmission between the two waveguides, but it is still very strong. Should the plasma frequency limit for this device be slightly increased, the transmission difference between ports 2 and 3 would likely increase at the expense of long-term performance of the plasma elements or the time scales upon which the device could be active. With the limit as it is at $f_p=7$ GHz, this device could be operated at very long time scales ($\sim$ hours) for many months as this is essentially the nominal operating condition specified by the bulb manufacturer. Driving the elements slightly harder would reduce the amount of time they could be active before runaway thermal effects begin to degrade performance on both the large and small time scales; though in order to reach catastrophic degradation the discharges have to be pushed quite far beyond the 7 GHz limit.

The final configurations shown in Figs. \ref{fig:straightguides} and \ref{fig:bentguides}, much like in the previous study \cite{RodriguezPRA}, appear to utilize the plasma elements in a much more nuanced way than a human-designed device would. Human-designed devices for a configuration like this would resemble a simple photonic crystal waveguide like that of Wang et al. in ref. \cite{guideppc}, where all of the discharges within the device are tuned to their maximum plasma density except for those along the path that the user would like the source to propagate, seeking to mimic a metallic waveguide. For any finite device with non-ideal elements, this more naive approach results in much more loss and very little signal at the correct exit, particularly in the bent waveguide case. The plasma density distribution obtained via inverse design, despite its strange structure which, in regions, mirrors the seemingly random structures that often arise from inverse design schemes \cite{piggott2015inversebinarized, molesky2018inverse}, has some plasma elements that are activated along the path of propagation of the source, even partially covering the entrance and exit which is entirely unintuitive. We suspect that the algorithm is finding an optimal balance between utilizing the reflective behavior of over-dense plasma and the refractive behavior of under-dense plasma to steer the light in a more effective way than simply attempting to mimic a metallic waveguide. It is also interesting to note how the devices create a gradually expanding and then contracting path within the device, similar to two horns directed at one another. This geometry is likely part of what reduces the insertion loss compared to the case when the plasma elements are not present. This demonstrates that inverse design allows us to arrive at novel, highly effective configurations that would not be formulated through conventional means.

\begin{figure}[htbp!]
\centering
\includegraphics[width=1\linewidth]{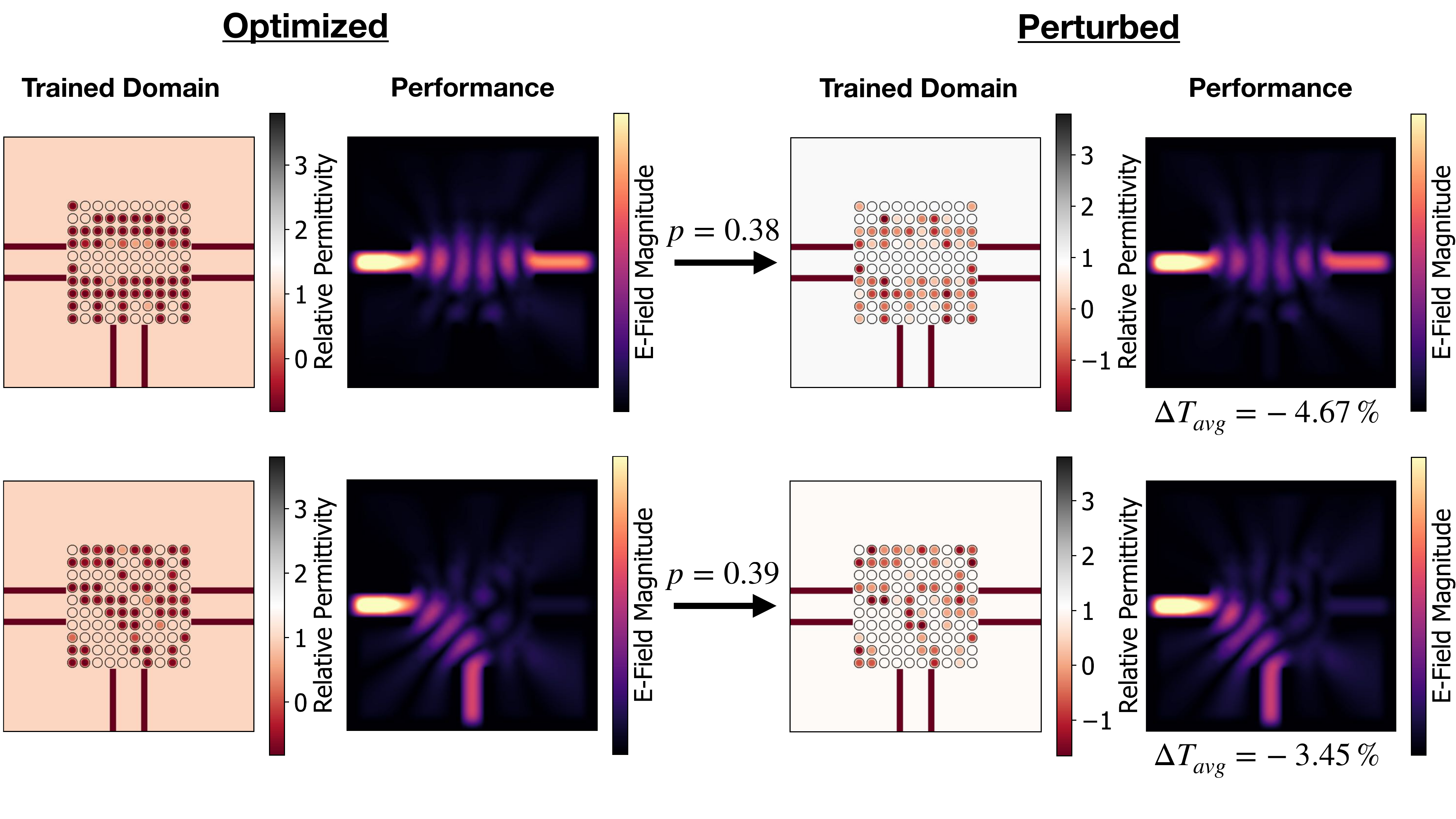}
\caption{Result for the waveguide stress test including the perturbed domain and field results along with average transmission drop in the correct output port across 10 perturbations with the denoted value of $p$.}
\label{fig:wvgpert}
\end{figure}

In fig. \ref{fig:wvgpert} we see the results of the sensitivity analysis for the best straight and bent waveguide configurations. In both cases we see that the optimal design is very robust to random perturbations of the device parameters as $p$ needs to be increased to $\geq 0.38$ to see an appreciable loss in functionality. Some of the rods in this case are so greatly altered that at the peak plasma density at the center of the plasma rods the permittivity drops as low as -1.5 where before the perturbations it was -0.7, and likewise many of the rods are nearly completely inactivated. The field plots show a very minor increase in loss to the surroundings and a drop in transmission at the corerct port, but even when 3 of the 10 perturbed devices failed, the average drop in transmission among the perturbed devices was only -4.67\% in the straight waveguide case and -3.45\% in the bent waveguide case. Keeping experimental error within $\sim 38\%$ is well within current capabilities and so we can assume that this particular non-ideal factor will not hinder experimental realization of these designs.

The results demonstrated in Figs. \ref{fig:straightguides}, \ref{fig:bentguides}, and \ref{fig:wvgpert} are reassuring. As mentioned before, simple waveguides such as these can serve as building blocks for more complicated devices, and strong performance like what we see here with very conservative operating conditions opens up many possibilities. It is also important here to remind the reader that unlike most devices produced via inverse design, the PMM geometry allows a single device to perform both functions as the plasma elements can be tuned electronically. As an example, quickly switching between the straight and bent propagation mode could enable the design of transistor-like optical switches that are experimentally feasible. The square lattice PMM configuration is capable of more than just waveguiding, however, as we will explore in the next subsection.

\subsection{Demultiplexer}
Next, we present a demultiplexer designed to discriminate between frequencies of $\tilde{\omega}_1 =0.25$ and $\tilde{\omega}_2=0.27$. Inverse design methods have been utilized in the past to create frequency demultiplexers \cite{piggott2015inversebinarized, PestourieMetasurfacesDemultiplex}, and the demultiplexer is a key component in many optical computing schemes \cite{XuONN}. The optimization objective for the demultiplexer is quite different from the waveguide case since the objective must take into account two separate simulations as the permittivity of the elements is frequency-dependent. With this in mind, the demultiplexer objective is:
\begin{widetext}
	\begin{equation*}
		\mathcal{L}_{mp} = \left(\int E_{\omega_1}\cdot E_{m=1}^*dl_{\omega_1\text{ exit}}\right)\left(\int E_{\omega_2}\cdot E_{m=1}^*dl_{\omega_2\text{ exit}}\right)-\int |E_{\omega_1}|^2dl_{\omega_2\text{ exit}}-\int |E_{\omega_2}|^2dl_{\omega_1\text{ exit}},
	\end{equation*}
\end{widetext}
where $E_{\omega_1}$ is the simulated field for the $\tilde{\omega}=0.25$ source and $E_{\omega_2}$ is the simulated field for the $\tilde{\omega}=0.27$ source. Since this objective is significantly more complex than the waveguide case, we allow the plasma frequency of the rods to increase up to $f_p=13$ GHz, an operating condition which can be maintained for time scales $\leq 10$ seconds at a time without damaging or otherwise altering the operation of the discharge bulbs. If operated in short ($\sim 1$-10 ms) pulses with a duty cycle that limits the proportion of time that the bulbs are active, this higher plasma density will not damage the discharges over much larger time scales (e.g. the pulsed operation affects the elements in a similar manner to the nominal operating conditions) and thus represents a very reasonable operating condition that could be sustained over a long device lifetime.
	
As an additional result of the significant jump in complexity from the waveguide case, it is not sufficient to simply take the optimal parameters from the idealized study \cite{RodriguezPRA} and take those as our initial parameters. Instead, we borrow from this idea and train the device parameters in several steps, starting with all the plasmas off and training a domain where the only non-ideal factor is the presence of the quartz envelopes which has an objective function with fewer local minima, allowing the algorithm to converge quickly on a configuration with strong performance. Then, once an optimal design was obtained for this simpler case, we introduce the non-uniform plasma density profile and train using the optimal parameters from the previous run. Finally, we add loss to the rods, train once again and arrive at a final trained domain. The parameter evolution along with the device performance is shown below in figure \ref{fig:demult}. While this design exhibits much more insertion loss and dissipation within the plasma elements, the difference in transmission between the correct and incorrect output ports is very large and the device performance is very similar to that achieved in ref. \cite{RodriguezPRA}.
	
\begin{figure}[htbp!]
	\centering
	\includegraphics[width=\linewidth]{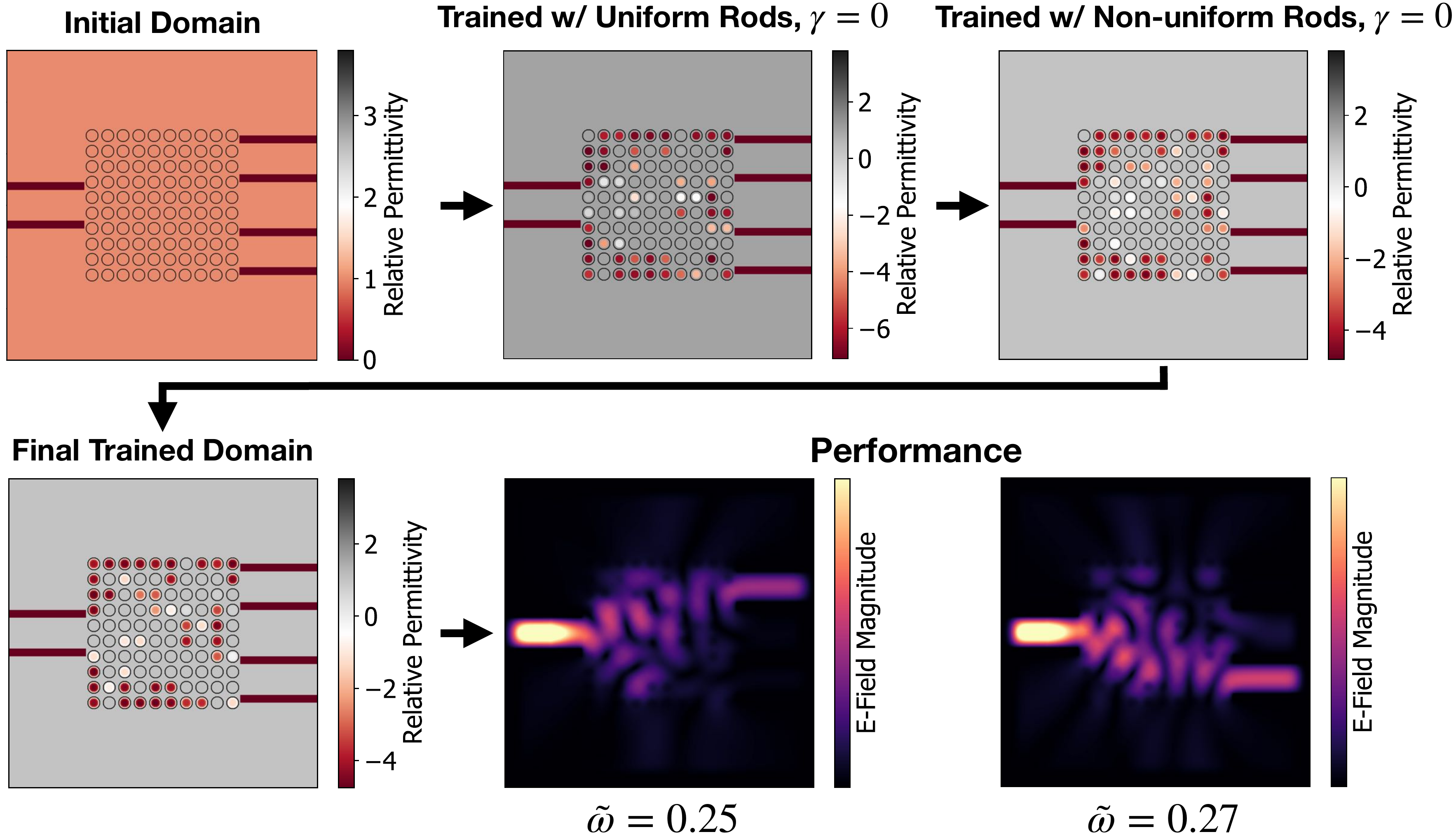}
	\caption{Evolution of the relative permittivity of demultiplexer simulation domain throughout the training stages as the plasma model complexity is increased, along with FDFD simulations showing final device performance at each operating frequency. The maximum plasma frequency among the elements in this case is about $\tilde{\omega}_p=0.8$ which translates to $f_p=13$ GHz.}
	\label{fig:demult}
\end{figure}
	
To conclude our analysis of the demultiplexer, we run our stress test procedure on the final design and find that in order to see 3 failures out of 10 perturbed cases, the perturbation parameter $p$ must be increased all the way to 0.31, which is very close to the values found in the waveguide sensitivity test above. The optimized and perturbed domain along with the degraded performance can be found below in figure \ref{fig:multpert}. We also note that since the degraded performance of 10\% or greater in 3 perturbations out of 10 is required for both frequencies, the average transmission drop across the 10 perturbations is higher. Even with this considered, the high $p$ needed to produce this level of degradation indicates that the plasma density distribution that we arrived at in this case is very robust to experimental noise/error.
	
\begin{figure}[htpb]
	\centering
	\includegraphics[width=1\linewidth]{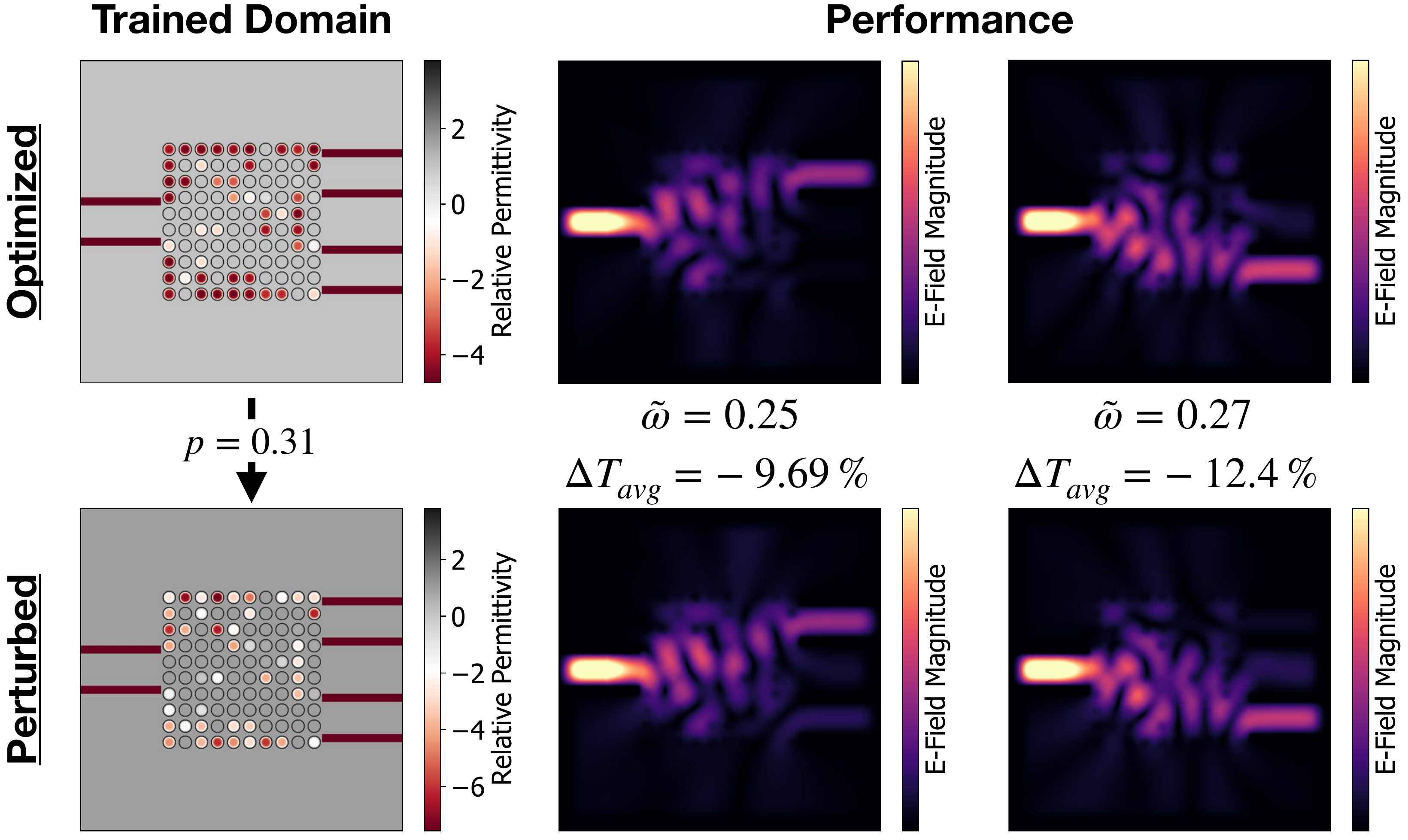}
	\caption{Result for the demultiplexer stress test including the perturbed domain and field results along with average transmission drop in the correct output ports across 10 perturbations, showing a significant degradation of performance when the perturbation parameter $p$ is increased to 0.31.}
	\label{fig:multpert}
\end{figure}

The results shown here are very compelling. For both source frequencies, the field intensity at the correct output waveguide is substantially higher than at the incorrect output (2-3 orders of magnitude higher as shown in fig. \ref{fig:Sparams}(b)), and the designs are extremely robust to random perturbation of the parameters. In addition, the insight gained from the optimization procedure for the demultiplexer objective in particular paves the way for both experimental realization of these and more complex devices, such as the logic gates presented in ref. \cite{RodriguezPRA}. If one wants to achieve a complicated device function, we have shown here that a good way of doing so is to optimize in stages while adding one non-ideal factor at a time, and then perhaps even perform \textit{in-situ} training of the device using the optimal parameters of the most realistic optimized device as the initial parameters. In the end, these results suggest that the outlook for experimental realization is very bright.

\section{Conclusion}
This work opens up many opportunities for future work, both in simulation and experiment. On the simulation front, with the improved plasma element model, new device factors could be introduced, such as different geometries (hexagonal lattices, different experimental parameters/bulb designs, larger arrays, etc.), variable operating frequencies, externally applied magnetic fields, and time-dependent behavior. For magnetized plasma in particular, the use of circularly-polarized waves gives the user access to a resonance that makes the plasma capable of achieving $\varepsilon>1$, making much more complex operations possible. Since Ceviche also allows for time-domain simulations, the non-linear response of the plasma elements to a high-power source could also be explored, eventually enabling the design of highly sophisticated devices like all-optical recurrent neural networks. In addition to all of this, the model for our plasma elements could be further improved via the use of kinetic particle-in-cell simulations for each bulb given an operating voltage, current, and fill pressure/composition. The particle density functions obtained from these codes could be used to provide a more accurate density profile than the polynomial profile used in this study, and more detailed collision modeling can be incorporated. With such an accurate model of our elements, one could likely easily take optimal parameters straight from the \textit{in-silico} inverse design algorithm and achieve high levels of performance in the lab. 

However, should experimental realization via the direct use of simulation results prove challenging, the nature of the PMM device, specifically the fact that its elements can be tuned continuously in real-time, allows for the training to be carried out fully \textit{in-situ} with plasma element currents adjusted while the gradient of the objective (transmission at a certain point, for example) is determined in real-time. This could be achieved by either carrying out the Ceviche forward-mode differentiation scheme using the experimental device in place of the simulation, or even a more complicated field measurement-based approach that would allow one to calculate the gradients through back-propagation \cite{insituInverseDesign}. Doing the device training \textit{in-situ} would automatically take all non-ideal factors into account. The design and construction of such a system will not be straightforward as we require programmatic control of $\sim100$ individual plasma discharge systems that can be smoothly varied in plasma density, but work has already begun on this front and is proceeding nicely.

In conclusion, we apply inverse design methods to create highly optimized and realistic two-dimensional PMM devices including directional waveguides and a demultiplexer for the TM polarization. Despite the consideration of several non-ideal factors such as the presence of quartz envelopes, collisionality/loss, non-uniform density profiles, and experimental error/noise, the devices all perform well and exhibit nuanced, complicated designs that would not be possible without inverse design methodology. A robust procedure for producing effective, experimentally realizable PMMs consisting of a gradual increase in model complexity is presented. The optimal parameters of each of the devices are experimentally feasible with existing equipment, enabling imminent realization of these designs. Further work is needed in both simulation and experiments to enable development of devices with more exotic objectives like boolean logic \cite{RodriguezPRA} as well as others with complicated EM wave propagation structures.

\begin{acknowledgments} 
This research is partially supported by the Air Force Office of Scientific Research through a Multi-University Research Initiative (MURI) with Dr. Mitat Birkan as Program Manager. J.A.R. acknowledges support by the U.S. Department of Energy, Office of Science, Office of Advanced Scientific Computing Research, Department of Energy Computational Science Graduate Fellowship under Award Number [DE-SC0019323].
\end{acknowledgments}

\bibliography{refs}

\end{document}